\documentclass[twocolumn]{aa}
\usepackage{graphicx}
\usepackage{txfonts}
\usepackage{amsfonts}
\usepackage{bm}
\def\mrho{\mbox{\boldmath $\rho$}}

\begin{document}                                                                                   

\newcommand{\bhat}{\mbox{${\bf\hat{b}}$}}
\newcommand{\ehat}{\mbox{${\bf\hat{e}}_1$}}

\title{Gyrokinetic electron acceleration in the force-free corona with anomalous resistivity} 
\titlerunning{Gyrokinetic electron acceleration}

\author{K. Arzner
          \inst{1}
          \and
          L. Vlahos\inst{2}
          }

\offprints{K. Arzner}

\institute{Paul Scherrer Institut, CH 5232 Villigen, Switzerland\\
              	\email{arzner@astro.phys.ethz.ch}
         \and
	 	Department of Physics, Aristotle University of Thessaloniki, 54124 Thessaloniki, Greece\\
             	\email{vlahos@astro.auth.gr}
             }

\date{Received 2. February 2006; accepted 27. March 2006}
	  
\abstract
{}
{We numerically explore electron acceleration and coronal heating by dissipative electric fields.}
{Electrons are traced in linear force-free magnetic fields extrapolated 
from SOHO/MDI magnetograms, endowed with anomalous resistivity ($\eta$) in localized
dissipation regions where the magnetic twist $\nabla \times \bhat$ exceeds a given threshold.
Associated with $\eta > 0$ is a parallel electric field ${\bf E} = \eta {\bf j}$ which can accelerate 
runaway electrons. In order to gain observational predictions we inject electrons inside the dissipation 
regions and follow them for several seconds in real time.}
{Precipitating electrons which leave the simulation 
system at height $z$ = 0 are associated with hard X rays, and electrons which escape at height $z$ $\sim$ 
3$\cdot 10^4$ km are associated with normal-drifting type IIIs at the local plasma frequency. A third, trapped, 
population is related to gyrosynchrotron emission. Time profiles and spectra of all three emissions are calculated, 
and their dependence on the geometric model parameters and on $\eta$ is explored. It is found that 
precipitation generally preceeds escape by fractions of a second, and that the electrons perform many visits to 
the dissipation regions before leaving the simulation system. The electrons impacting $z$ = 0 reach
higher energies than the escaping ones, and non-Maxwellian tails are observed at energies above the
largest potential drop across a single dissipation region. Impact maps at $z$ = 0 show a tendency
of the electrons to arrive at the borders of sunspots of one polarity.}
{Although the magnetograms used
here belong to non-flaring times, so that the simulations refer to nanoflares and
`quiescent' coronal heating, it is conjectured that the same process, on a larger scale, 
is responsible for solar flares.}


\maketitle

\section{Introduction}

Observations of radio waves and hard X-rays (HXR) from solar flares allow the study of the
acceleration and propagation of high-energy electrons which are responsible for both
types of emission. The picture emerging from the observations is generally complicated
(Benz et al. \cite{benz05}). While HXR (bremsstrahlung) emission is often associated with
metric type III radio groups (Aschwanden et al. \cite{aschwanden95}), the timing 
of individual type III's and HXR fine structure is erratic. Metric type III's are presumably
caused by electron beams exciting Langmuir waves, which then couple to electromagnetic
(observable) modes. Sometimes, there is 
perfect agreement between type III onsets and HXR maxima, but in other 
cases there is no obvious peak-to-peak correlation, or one of the emissions is 
absent altogether. As a trend, the type III's onsets were found to be delayed by 
fractions of a second against the HXR fine structures (Aschwanden et al. \cite{aschwanden92}; 
Arzner \& Benz \cite{arzner05a}), although the type III frequency drift may assist
such delays in cases where the type III onset could not properly be resolved.
At millimeter and decimeter wavelengths, synchrotron emission (e.g., Gim\'enez et al. \cite{gimenez05}), 
decimetric radio continuum and decimetric pulsations (Saint-Hilaire \& Benz \cite{sainthilaire03}), and 
decimetric spikes (G\"udel et al. \cite{guedel91}) have also been found to be associated with HXR, 
roughly in the above decreasing order of association probability. These types of decimetric 
and millimetric radio emission are related to magnetically trapped electrons, where loss-cone 
instabilities are believed to be responsible for temporal structures (Kuijpers \& Slottje \cite{kuijpers76},
Aschwanden \& Benz \cite{aschwanden88}, Fleishman \& Melnikov \cite{fleishman98}).

The variety of observed behaviour accounts, on the one hand, for the non-linear (coherent) 
radio emission processes. On the other hand, it also reflects the geometric complexity of the active 
regions, which results from the nonlinear dynamics in the convection zone. In fact, the complex
behaviour and the complex geometry are likely to be causes and effects of each other. Including a 
realistic amount of geometric complexity into a numerical model of solar flares was a major motivation 
for the present work.

Apart of geometrical aspects, the modeling of solar flares requires the specification of a physical acceleration 
mechanism. We shall assume here, as a working hypothesis, that acceleration is caused by DC parallel electric fields due
to anomalous resistivity. Such fields are capable to accelerate high-energy electrons for which 
the electric force is no longer counterbalanced by the collisional drag (Dreicer \cite{dreicer60}). 
It should be pointed out that runaway acceleration applies only to a small fraction of electrons in the
high-energy tail of the electron distribution, and that the majority of electrons is (Joule) heated 
rather than accelerated to superthermal energies. The electric fields envisaged above are of
macroscopic and dissipative nature, and mark the irreversible release of 
non-potential magnetic energy. We thus follow a scenario, proposed by Parker (\cite{parker72},\cite{parker83},\cite{parker93}),
where the random photospheric footpoint motion twists and shears the coronal magnetic loops such
as to develop tangential discontinuities. This happens intermittently and
in multiple localized regions throughout the solar corona, and does not generally imply
loop instability or global re-structuring of the solar corona. Parker's idea has initiated many investigations
on coronal heating and -eruptions (e.g., Schumacher \& Kliem \cite{schumacher96}; T\"or\"ok \& Kliem \cite{torok01}; 
Gudiksen \& Nordlund \cite{gudiksen02}), and has significantly improved the understanding of the global flaring process.
In our view, a strong but heuristic argument for adopting dissipative (rather than conservative) electric fields
as responsible for particle acceleration in solar flares is the intermittent and violent nature of flares,
hinting to a catastrophic process which could not be cast in a Hamiltonian formalism.
However, conservative electric fields, as used in most investigations on
stochastic acceleration (e.g. Karimabadi et al. \cite{karimabadi87}; Schlickeiser \cite{schlickeiser03}),
may also act as particle accelerators.

In the present study, we adopt a test
particle approach similar as in the simulations of Matthaeus \& Lamkin (\cite{matthaeus89}),
Dmitruk et al. (\cite{dmitruk03,dmitruk04}), Arzner \& Vlahos (\cite{arzner04}), and Arzner et al. (\cite{arzner06}), 
but with two important modifications. First, the electromagnetic force fields
are not taken from MHD turbulence simulations or random-phase turbulence proxies, but
from observed magnetograms. Secondly, in order to span the many orders of magnitude between 
the electron Larmor radius and the size of magnetic loops, we allow for
gyrokinetic motion, keeping track of the gyro phase in an approximate way. This
technique enables us to follow the electrons over several seconds in real time (several $10^8$ gyro times), 
thus reaching the time scales on which solar flares are observed. In the simulations, 
runaway electrons are injected at $t$ = 0 inside the localized dissipation regions,
and followed numerically along (and sometimes across) the magnetic field lines.
Observational predictions for HXR and radio waves are obtained as the electrons impinge
the chromosphere, get trapped, or escape to the higher corona.

The article is organized as follows. Section 2 describes the construction of the coronal field; 
Section 3, the particle orbits; Section 4, the numerical results and observational predictions. These are then
summarized and discussed in Section 5.

\section{Coronal electromagnetic fields}

Our domain is a slab of height $H$, bounded at $z$ = 0 by the photosphere, and filled with 
time-independent force-free (e.g., Gary \cite{gary89}) magnetic fields. We do not ask here for a perfect field 
reconstruction, but for a generic configuration which is compatible with the observation. Accordingly, we use an
elementary linear force-free extrapolation from the normal photospheric magnetic field. This
represents a local approximation at best (Wheatland \cite{wheatland99}), and we have a slab thickness 
$H$ of not more than a few $10^4$ km in mind when fixing the physical scaling. (The linear force-free
assumption predicts loops flaring with height, in contrast to the slender high-ranging loops observed by TRACE.)
Further characteristic length and time scales are compiled in Table \ref{notation_tab}.

\begin{table}[h]
\begin{tabular}{ccc}
Symbol		& Meaning		 & Typical value \\\hline
$\Omega^{-1}$	& particle time unit	& $10^{-9}$ s \\
$l_0$		& particle length unit	& 0.3 m	\\
$l$		& magnetogram resolution& $660$ km \\
$L$		& magnetogram size	& $1.7 \cdot 10^5$ km \\
$H$		& slab height		& $3 \cdot 10^4$ km \\
$1/\alpha$	& force-free scale	& $2 \cdot 10^4$ km \\
$1/u_c$		& critical twist scale	& $3 \cdot 10^3$ km
\end{tabular}
\caption{\label{notation_tab}Notation and physical scaling.}
\end{table}

\subsection{Construction of the force-free magnetic field}

The force-free condition requires that $\nabla \times {\bf B} = \alpha {\bf B}$, where $\alpha$ is assumed
to be constant (`linear' force-free field). Our construction of the force-free magnetic field follows the lines 
of Alissandrakis (\cite{alissandrakis81}), with modifications concerning the selection and number of Fourier modes. 
In order to avoid large but passive 3-dimensional arrays, the strategy is to 
locally compute ${\bf B}({\bf x})$ from a restricted number of Fourier modes 
${\bf b} ({\bf k}) \, e^{i {\bf k} \cdot {\bf x}}$. In terms of these, the force-free condition becomes
$i {\bf k} \times {\bf b} ({\bf k}) = \alpha {\bf b} ({\bf k})$. This equation has a non-trivial solution only if 
$|{\bf k}|^2 = \alpha^2$, which we write in the form

\begin{equation}
k_z = \pm \sqrt{\alpha^2 - k_\perp^2} \;\;\;\; \mbox{where} \;\;\; k_\perp^2 = k_x^2 - k_y^2 \, , 
\label{disprel}
\end{equation}
 
and if ${\bf b}({\bf k})$ is proportional to

\begin{equation}
(- k_x k_z + i \alpha k_y , \; - k_y k_z - i \alpha k_x, \;  k_x^2 + k_y^2 )  \, .
\label{evects}
\end{equation}

See MacLeod (\cite{macleod95}) for a deeper discussion of the curl eigenfunctions.
We assume $k_x$ and $k_y$ to be real, so that the magnetic field in the ($x,y$)-plane is bounded,
while $k_z$ may be complex with positive imaginary part. Thus modes decaying or oscillating
with height are permitted, and we shall use a small admixture of oscillating modes to transport structures
from the chromosphere to larger heights. From a physics point of view this procedure is justified by 
noticing that the outstreaming solar wind (in a 1d slab geometry) does not require ${\bf B}(z)$ to vanish as 
$z \to \infty$.
\begin{figure}[ht]
\centerline{
\includegraphics[width=0.25\textwidth]{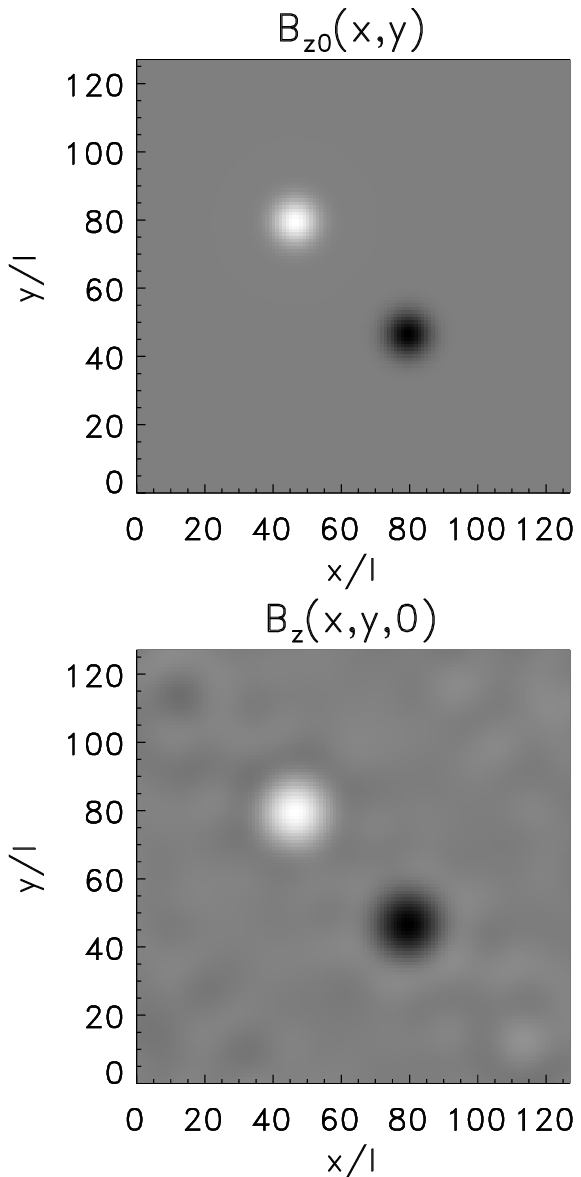}
\includegraphics[width=0.25\textwidth]{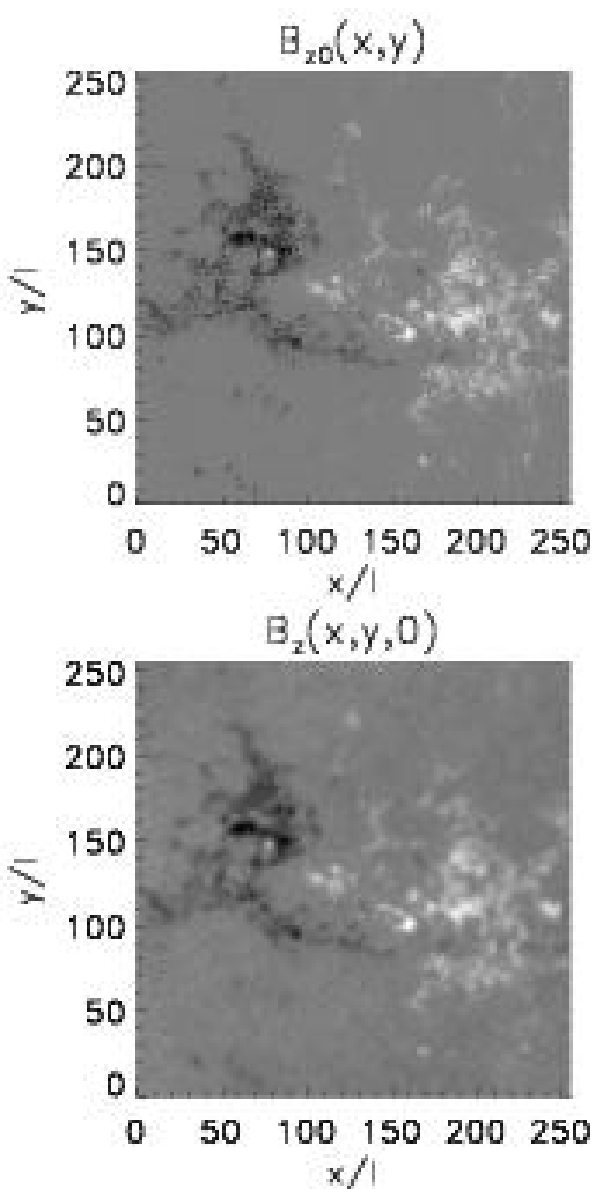}}
\caption{Boundary fields (top) and their sparse-Fourier representations
(bottom) of a bipolar configuration (left) and a SOHO/MDI magnetogram (right)
recorded on August 17, 2002. 10:40 UT. The lower left corner of the magnetogram
corresponds to $(78'',-36'')$ in heliocentric coordinates, and the scale is $l$ = 0.94''.}
\label{b0_fig}
\end{figure}
\begin{figure}[h]
\centerline{
\includegraphics[width=0.25\textwidth]{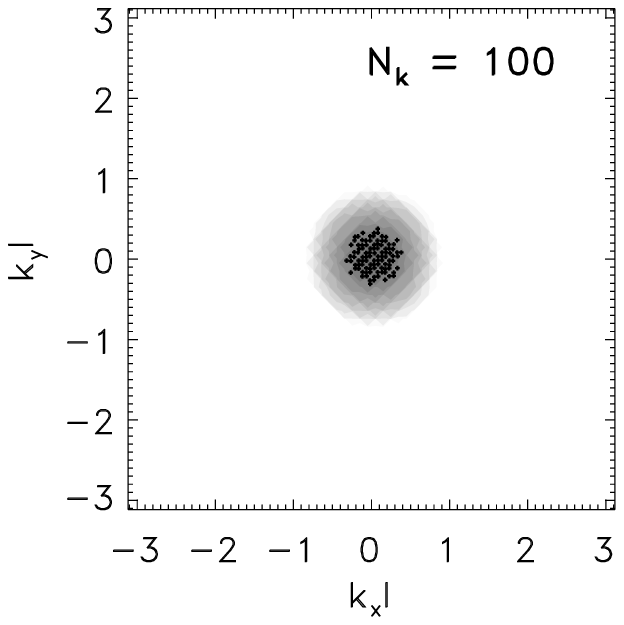}
\includegraphics[width=0.25\textwidth]{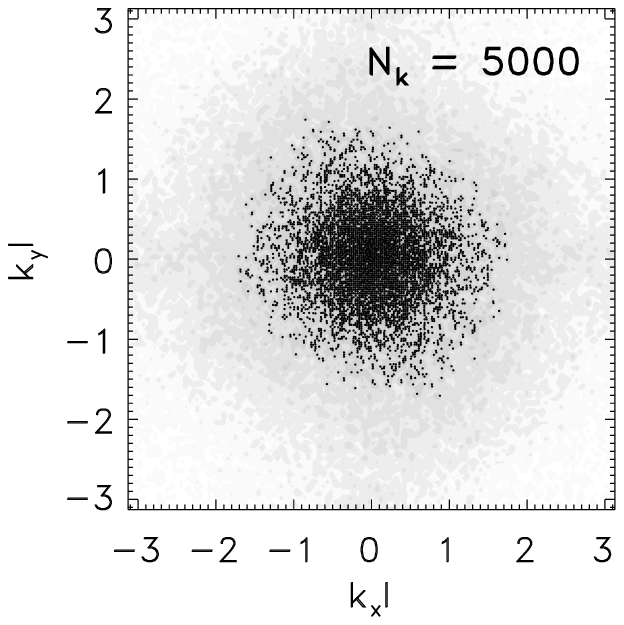}}
\caption{Power spectral density (grayscale) of the boundary fields of Fig. \protect\ref{b0_fig} (top). 
Left: bipolar configuration; right: SOHO/MDI magnetogram. The sparse-Fourier components are marked by dots.}
\label{s0_fig}
\end{figure}
Furthermore, different wave vectors are not rational multiples of each other, so that the magnetic field 
extends in a non-periodic, generic, way across the $(x,y)$-plane.
The irregular spacing of the Fourier components also avoids
aliasing artifacts, and their small number allows a non-expensive evaluation
of the magnetic field even if no fast Fourier transform is available.
However, the restriction to few Fourier components makes an exact matching of the
boundary conditions impossible. Instead, we require agreement in a least square sense only, and
minimize the mean-square deviation of the sparse-Fourier field $B_z(x,y,0)$ and the photospheric
boundary field $B_{z0}(x,y)$ in the square $L \times L$. Setting 

\begin{equation}
{\bf B}({\bf x}) = {\rm Re} \sum_{\bf k} C_{\bf k} {\bf b}({\bf k} )e^{i {\bf k}  \cdot {\bf x}} \, ,
\label{B}
\end{equation}

normalizing the eigenvectors such that $|b_z| = 1$, and using the (approximate) 
orthogonality of the harmonic functions over $L \times L$, the coefficients $C({\bf k})$ are given by

\begin{equation}
C({\bf k}) \simeq \frac{const.}{N_k} \int_{L \times L} \hspace{-2mm} dx \, dy \; b_z^*({\bf k}) e^{-ik_x x - ik_y y} B_{z0}(x,y) \, .
\label{Cn}
\end{equation}

The boundary field $B_{z0}(x,y)$ is given on a square lattice (cell size $l^2$), and the integral (\ref{Cn}) is computed numerically.
`Complementary' solutions of the homogeneous boundary value problem (Chiu \& Hilton \cite{chiu77}; Petrie \& Lothian \cite{petrie03}) 
are disregarded. The sparse-Fourier wave vectors $(k_x,k_y)$ are selected by the following procedure. First, the fast Fourier transform of
the discrete boundary field is computed and those $N_k$ (regularly spaced) wave vectors are chosen which have largest 
power spectral density. Then, a random perturbation $(\Delta k_x, \Delta k_y) \la \pi / l$ is added in 
order to break the exact periodicity. The eigenvectors (Eq. \ref{evects}) automatically ensure that 
$\nabla \cdot {\bf B} = 0$.

A test case with two Gaussian footpoints of opposite polarity is shown in Fig. \ref{b0_fig} 
(left column), with the true boundary field $B_{z0}(x,y)$ presented in the top left panel and its sparse-Fourier version $B_z(x,y,0)$
presented in the bottom left panel. The sparse-Fourier version contains $N_k$ = 100 components only. This low number 
already provides a good approximation (correlation coefficient 0.94), because the Fourier spectrum (Fig. \ref{s0_fig} left) 
is concentrated at the origin. When real data are used, a larger number of Fourier components is needed. This is 
illustrated in Figure \ref{b0_fig} (right column), showing a SOHO/MDI magnetogram (top right) and its sparse-Fourier 
representation (bottom right) using 5000 Fourier components (Fig. \ref{s0_fig} right). The magnetogram is located close to 
the center of the solar disc, so that projection effects are negligible. The magnetogram resolution is $l$ = 660 km (rescaled from 
the original SOHO/MDI resolution), and the correlation coefficient between the magnetogram and its sparse-Fourier
representation is 0.93. The magnetogram belongs to a non-flaring configuration and therefore the force-free extrapolation should be a reasonable 
approximation. Although the smallest scales are not well represented by the sparse-Fourier
approximation, they rapidly decay with height (Eq. \ref{disprel}), so that the approximation becomes
better as $z$ increases. This fact is used to save computation time by restricting the field computation
to components with $|e^{i k_z z}| > 10^{-3}$.

\begin{figure}[ht]
\centering
\includegraphics[width=0.48\textwidth,height=0.43\textwidth]{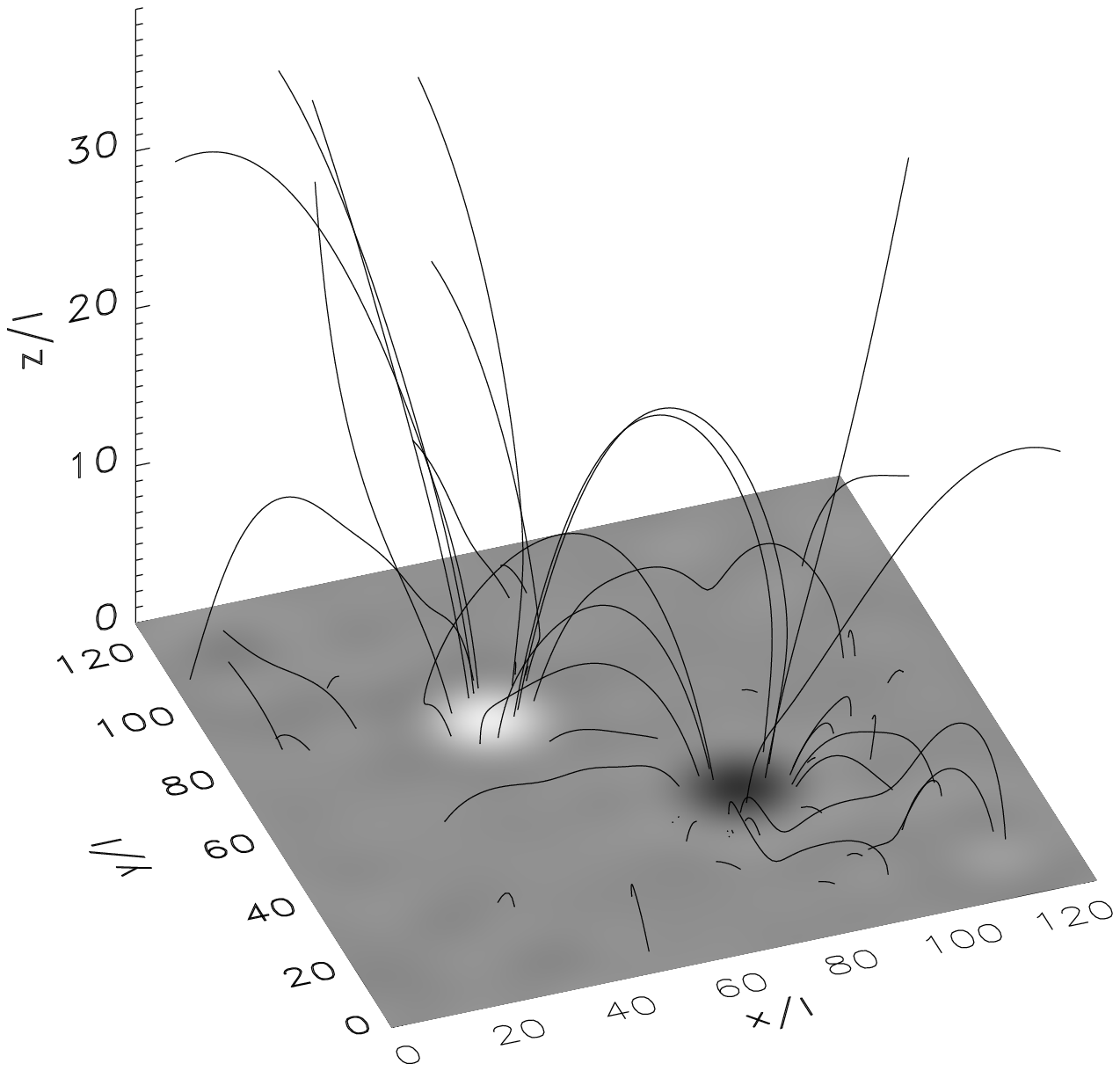}\\[-18mm]
\includegraphics[width=0.48\textwidth,height=0.43\textwidth]{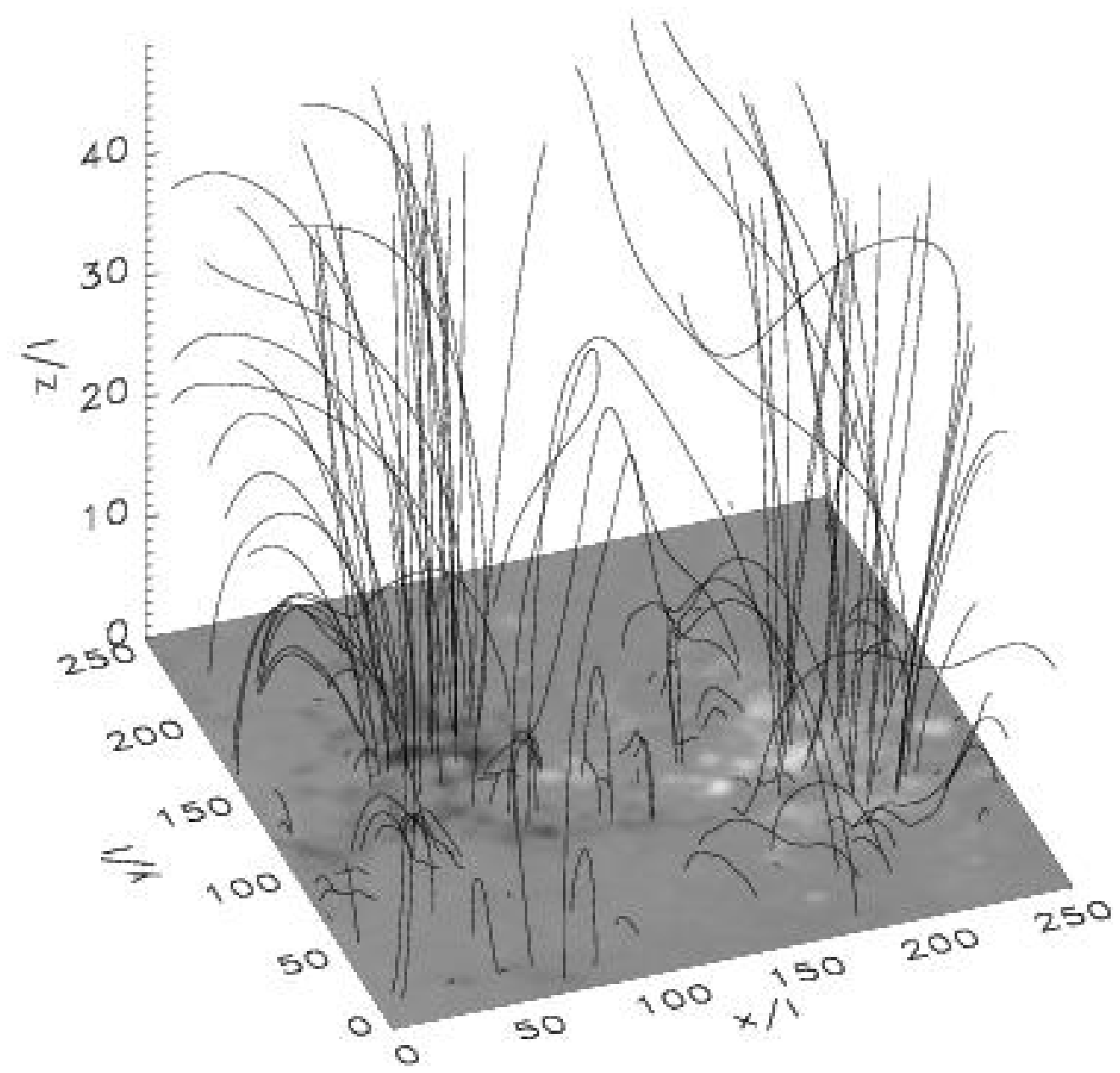}
\caption{Sparse-Fourier force-free magnetic field lines extrapolated from bipolar (top) and magnetogram (bottom) boundary data.}
\label{fieldlines_fig}
\end{figure}

\subsection{\label{field_sect}Properties of the magnetic field}

Figure \ref{fieldlines_fig} displays the force-free magnetic fieldlines of the bipolar test case (top) and the SOHO/MDI 
magnetogram (bottom). The $\alpha$ parameter is such that $\alpha l$ = 0.03, corresponding to a force-free scale $1/\alpha$ of 
about 2$\cdot 10^4$ km. The field lines start out at $z$ = 0 at random with density proportional to $|{\bf B}|$. Several field 
line integration schemes have been tested, and the force-free condition was also verified using 
finite 3D differences. Most field lines connect the two poles of opposite polarity, but not all, because the sparse-Fourier field 
does not vanish outside the poles, nor outside $L \times L$.

The absolute scaling of ${\bf B}$ is constrained by the SOHO/MDI data which give, for Figure \ref{fieldlines_fig} (bottom),
a longitudinal photospheric rms field $\sqrt{\langle B_{z0}(x,y)^2\rangle}$ = 50G, with excursions to $\pm 400$G at the footpoints. 
These values are typical for active regions; the average over the quiet solar surface
is smaller ($\sim$1G, see Kotov \cite{kotov02}). The mean-square magnetic field strength 
is related to the Fourier amplitudes by $\langle B_x(x,y,0)^2 \rangle =
\frac{1}{2} \sum_{\bf k} |b_x({\bf k})|^2$ and similar relations for $B_y$ and $B_z$.
 
\begin{figure}
\includegraphics[width=0.48\textwidth,height=0.28\textwidth]{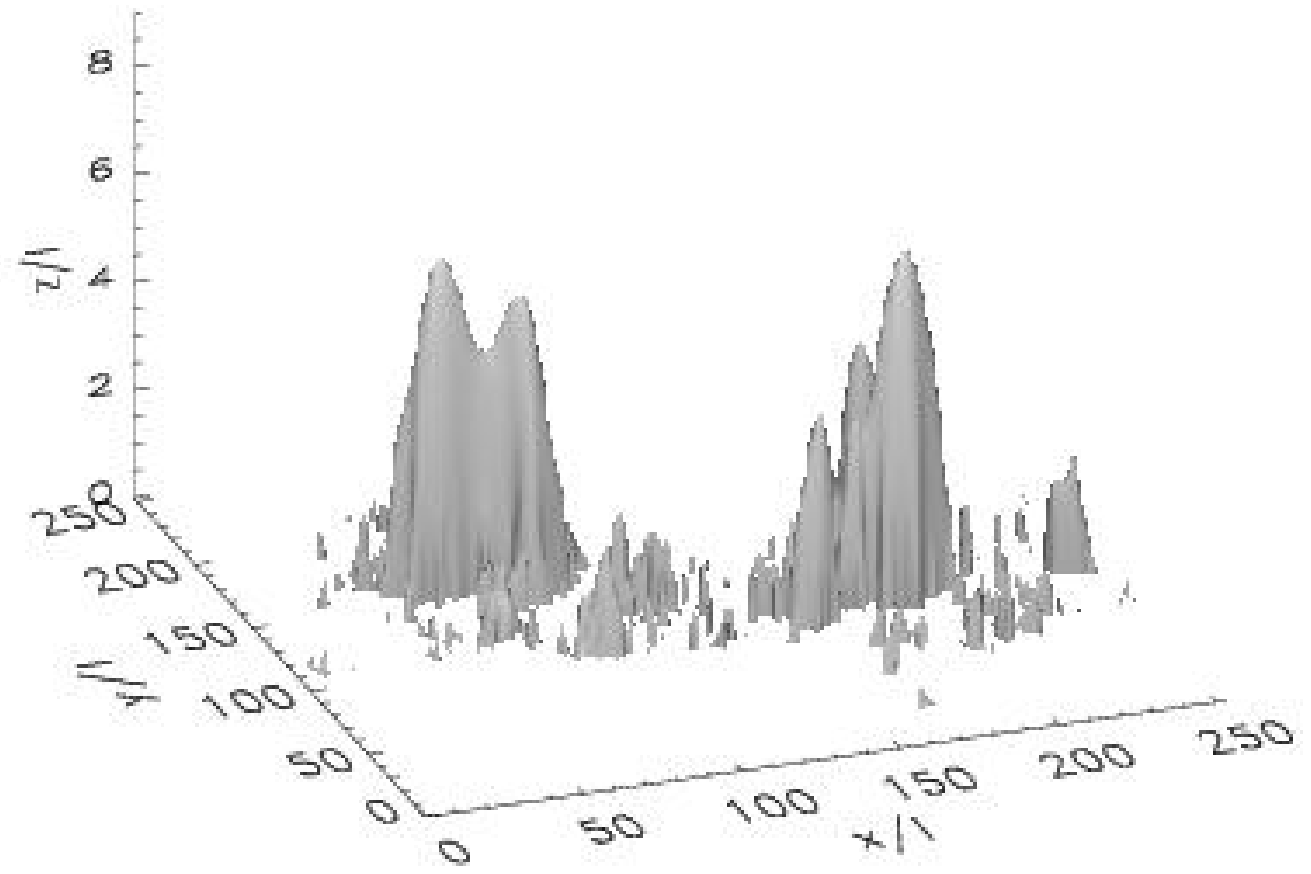}
\includegraphics[width=0.48\textwidth,height=0.28\textwidth]{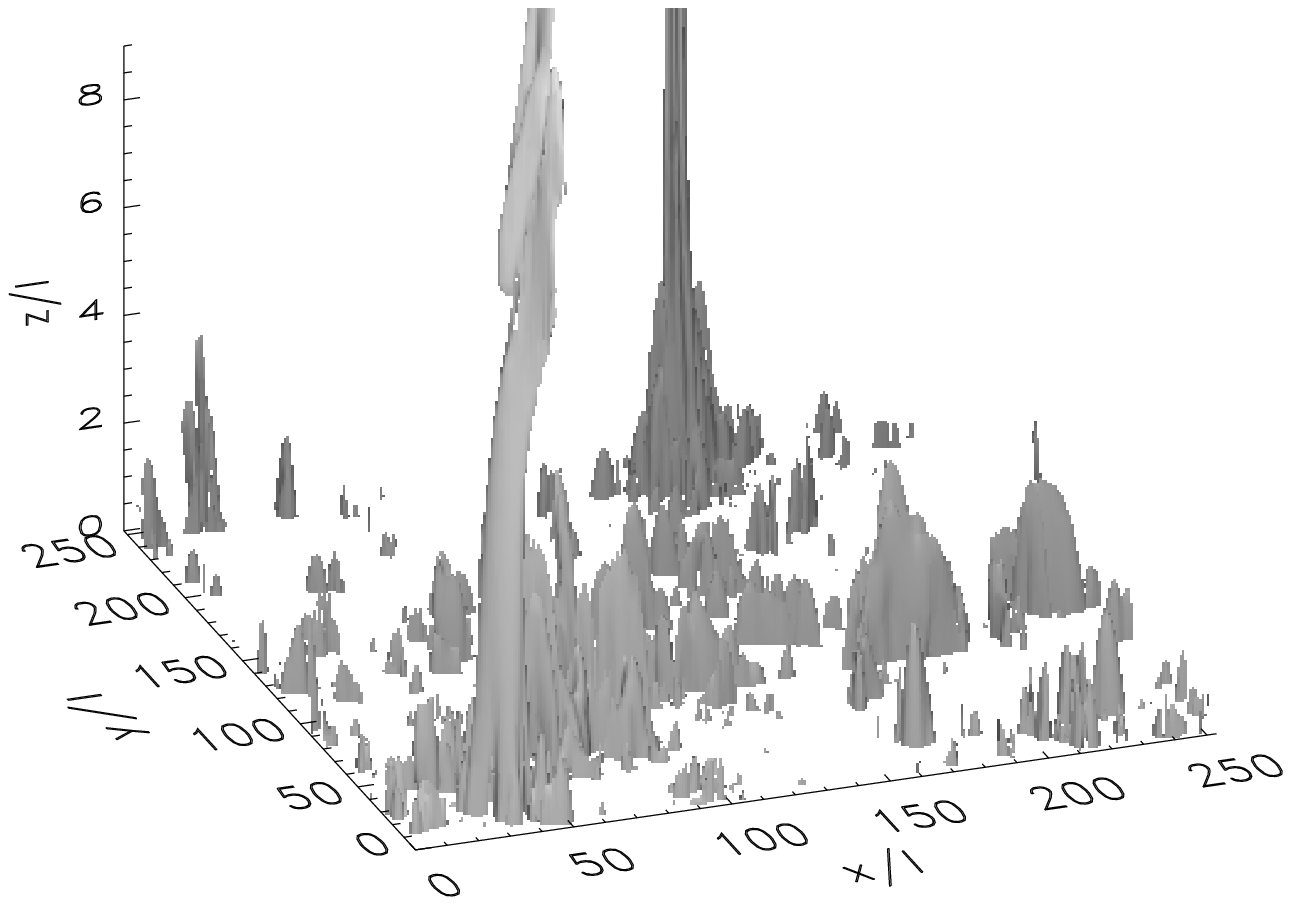}
\caption{Different definitions of dissipation regions. Top: current threshold. Bottom: magnetic twist threshold. Both
panels refer to the magnetogram of Fig. \protect\ref{fieldlines_fig} (bottom). It is the twist threshold which is used in the 
sequel of this paper.}
\label{ju_fig}
\end{figure}

\subsection{Electric field}

Since the configuration is assumed to be time-independent, the electric field has only an Ohmic
component ${\bf E} = \eta {\bf j}$. The (scalar) resistivity $\eta$ is either zero or anomalous, i.e.,
much larger than the collisional value $\eta_{\rm Spitzer}$ (Helander \cite{helander02}). Regions with $\eta \gg \eta_{\rm Spitzer}$
are called here `dissipation regions'.
Owing to the force-free condition $\nabla \times {\bf B} = \alpha {\bf B}$, the electric field ${\bf E} = \eta \alpha {\bf B}$ 
is purely parallel.

Different criteria for the occurrence of anomalous resistivity have been proposed, relying on
linear micro-instabilities or heuristic arguments. A well-known argument limits the electric
current to $j_c$ = $e n c_s$ where $c_s$ is the speed of sound (Papadopoulos \cite{papadopoulos80}); an excess $|\nabla \times {\bf B}| > j_c$ 
is then argued to produce shocks and dissipation in the current-carrying electron fluid. The resulting dissipation region 
$|{\bf j}| > j_c$ is depicted in Fig. \ref{ju_fig} (top) for the configuration of Fig. \ref{fieldlines_fig} (bottom). 
As the force-free current is proportional to the magnetic field, the dissipation regions simply delineate the magnetic field strength.

Another approach invokes the twist or shear of magnetic field lines rather than the electric current density. 
The physical motivation for this stems from the observation that regions of larger magnetic field should also be able to
guide larger currents before disruption. Thus the automatic increase of $j_c$ with increasing $|{\bf B}|$ should be
compensated, which is achieved by considering the quantity

\begin{equation}
\bhat = \frac{{\bf B}}{|{\bf B}|}
\label{bhat}
\end{equation}

rather than ${\bf B}$. The criterion for the occurrence of {\bf anomalous} resistivity is then defined by

\begin{equation}
|\nabla \times \bhat| < u_c \, .
\label{crit_cond}
\end{equation}

The resulting dissipation regions are shown in Fig. \ref{ju_fig} (bottom); the top and
bottom panels represent equal dissipation volumes.

It follows from the force-free condition that $\bhat \cdot \nabla \times \bhat = \alpha$. Thus, for
constant $\alpha$, the field-aligned component of the magnetic twist $\nabla \times \bhat$ 
is constant throughout the whole volume. This implies, in particular, that $|\nabla \times \bhat|$ can never 
be smaller than $\alpha$, and that the threshold $u_c$ must be larger than $\alpha$. In Fig. \ref{ju_fig}
(bottom), $u_c/\alpha = 6.6$. Since small scales with $k_\perp > \alpha$ decay exponentially with height, 
the steep gradients, and thus the dissipation regions, tend to accumulate at low altitudes. This is in agreement with
the observation that the coronal heating function is localized within a height range $\la 10^4$ km 
(Aschwanden \cite{aschwanden01}). The dissipation regions are thus
not uniformly distributed in height, and the dissipative volume $V_d$ can be characterized by a column height 

\begin{equation}
h_d = V_d/L^2
\label{hd}
\end{equation}

rather than by a volume fraction referring to the full numerical domain $L^2 H$. In
the case of Fig. \ref{ju_fig} one has $h_d$ = 0.19$l$.

The condition (\ref{crit_cond}) is local.
Non-local versions of Eq. (\ref{crit_cond}) have been used by Georgoulis \& Vlahos (\cite{georgoulis98})
along pairs of magnetic field lines, and by Vlahos \& Georgoulis (\cite{vlahosgeorgoulis04}) on a
discrete lattice. We use here Eq. (\ref{crit_cond}), since it can be evaluated
at the particle position and thus simplifies the simulation. The threshold $u_c$ is
chosen such that $u_c l$ = 0.2. 

\section{Particle orbits}

\subsection{Dimensionless units and physical scaling}

The system of units used in this article is similar as in Arzner \& Vlahos (\cite{arzner04}).
Time is measured in units of $\Omega^{-1}$ where $\Omega^2 = \frac{1}{2} \sum_{\bf k} |{\bf b}({\bf k})|^2$ is the non-relativistic 
mean-square gyro frequency at $z$ = 0. Velocity is measured in units of the speed of light; energy is measured in units of 
$m c^2$; momentum is measured in units of $mc$; distance is measured in units of $l_0 = c \Omega^{-1}$. We take 
$\langle |{\bf B}(x,y,0)|^2 \rangle^{1/2}$ = 50G (see Sect. \ref{field_sect}) to fix the absolute scaling, so that 
the electron cyclotron frequency is 140 MHz, the time unit is $\Omega^{-1} = 10^{-9}$s, and the length unit is $l_0 =$ 0.3m. 
The electric field is measured in units of $cB$, and the resistivity $\eta$ is measured in units of $c^2\Omega^{-1}$.

\subsection{Exact orbits}

In the above unit system, the exact particle equations of motion read

\begin{eqnarray}
\frac{d {\bf x}}{dt} & = & {\bf v} \label{dxdt} \\
\gamma \frac{d {\bf v}}{dt} & = & {\bf v} \times {\bf B} + {\bf E} - ({\bf v} \cdot {\bf E}) {\bf v} \label{dvdt} \, 
\end{eqnarray}

where ${\bf E} = \eta \nabla \times {\bf B}$ and $\gamma = 1/\sqrt{1-v^2}$ is the Lorentz factor. Equations (\ref{dxdt}) and (\ref{dvdt}) 
are integrated by a traditional Cash-Karp/Runge-Kutta method (Press et al. \cite{press98}) with adaptive
time step. The sparse-Fourier fields are computed at a lower rate, together with their Jacobians, from which a 
linear extrapolation to the actual particle position is made. The time step is chosen such that both the Cash-Karp error 
and the field extrapolation error remain within given bounds (Appendix \ref{numerics_appendix}).

\subsection{Gyrokinetic approximation} 

In the magnetic fields considered here, ranging from a few to a few hundred Gauss, and for typical kinetic energies between 
1 keV and 1 MeV, the electron Larmor radius $\gamma v_\perp /\Omega$ varies from centimeters to hundreds of meters, 
and is therefore small compared to the resolution of the magnetic field ($l \sim 10^3$ km).
The motion is thus almost always adiabatic, with possible exceptions at critical
points of the magnetic field, and inside the dissipation regions. This 
suggests a gyrokinetic approximation, with the possibility to switch to exact orbits if necessary.

We base our guiding center mover on a relativistic (Brizard \cite{brizard99}) version of the gyrokinetic Equations of 
Littlejohn (\cite{littlejohn81,littlejohn83}). The variables considered are $({\bf X}, p_\parallel, \Theta, \mu)$, where 
${\bf X}$ is the guiding center position, $p_\parallel = \gamma v_\parallel $ is the momentum parallel to the 
magnetic field, $\Theta$ is the gyro phase, and $\mu = \frac{1}{2} \gamma^2 v_\perp^2/B$ is the
magnetic moment, which is a motion integral in the present approximation. The coordinates $({\bf X},p_\parallel)$ 
evolve according to

\begin{eqnarray}
\frac{d {\bf X}}{dt} 	& = & \frac{1}{B_\parallel^*} \Big( \frac{p_\parallel}{\gamma} {\bf B}^* +  \frac{\mu}{\gamma} {\bf \hat{b}} \times 
				\nabla B - \bhat \times {\bf E}^* \Big) \label{dXdt} \\
\frac{d p_\parallel}{dt}& = & - \frac{1}{B_\parallel^*} {\bf B}^* \cdot \Big( \frac{\mu}{\gamma} \nabla B - {\bf E}^* \Big) \label{dUdt}
\end{eqnarray}

where $\gamma = \sqrt{1 + p_\parallel^2 + 2 \mu B}$ and

\begin{eqnarray}
{\bf B}^*	& = & {\bf B} + p_\parallel \nabla \times {\bf \hat{b}} \label{B*} \\
{\bf E}^*	& = & {\bf E} - p_\parallel \frac{\partial {\bf \hat{b}}}{\partial t} \label{E*}
\end{eqnarray}

with $B_\parallel^* = {\bf B}^* \cdot {\bf \hat{b}}$.
For our time-independent force-free magnetic field, ${\bf E}^*$ equals ${\bf E}$ and the last term
in Eq. (\ref{dXdt}) vanishes.

In order to switch back to exact orbit integration we need to keep track of the gyro phase. 
In gyrokinetic approximations, the gyro phase is not unambiguously defined but subject to gauge freedom.
Accordingly, various definitions have been proposed. We follow here Littlejohn (\cite{littlejohn88}). Let ${\bf \hat{e}}$
be a unit vector in direction of the initial ($t$=0) Larmor radius. The vector $\ehat$ is perpendicular
to the direction ${\bf \hat{b}}$ of the local magnetic field, and a local cartesian triad is completed by setting
${\bf \hat{e}}_2 = \ehat \times {\bf \hat{b}}$. The task is to follow the unit vector ${\bf \hat{e}}_1$
along the gyro center, introducing a minimum of twist. This is achieved by solving the parallel transport
(Fermi-Walker type) Equation

\begin{equation}
\frac{d \ehat}{ds} = - \bhat \Big(  \frac{d \bhat}{ds} \cdot \ehat \Big)
\label{duds}
\end{equation}

where $s$ is the distance along the gyrocenter orbit. The gyro phase, with respect
to the direction $\ehat$, is then given by $\Theta = \int_0^t \gamma^{-1} |{\bf B}({\bf X}(t'))| \, dt'$,
and the exact orbit parameters (${\bf x},{\bf v}$) can be retrieved from $({\bf X}, v_\parallel, \mu, \Theta, \ehat)$. 
More details and the benchmarking of our numerical approach are described in Appendix \ref{numerics_appendix}.

\section{Simulation results and observational predictions}

\begin{figure}[ht]
\centerline{\includegraphics[width=0.48\textwidth]{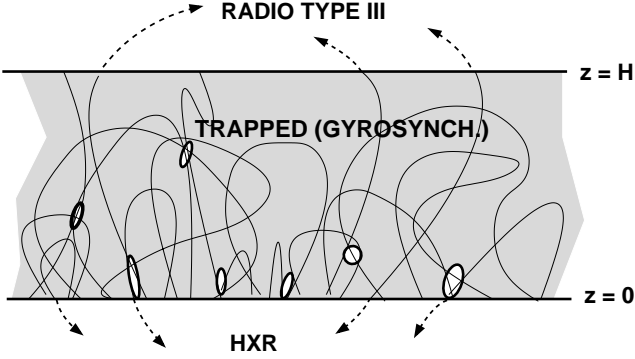}}
\caption{Schematic allocation of different emissions from high-energy electrons.
Precipitating electrons are associated with HXRs; escaping 
and trapped electrons are associated with radio emission.
Black lines symbolize magnetic field lines; boldface ellipses 
symbolize dissipation regions.}
\label{scheme_fig}
\end{figure}

The slab-shaped simulation domain suggests a simple scheme of observational diagnostics,
which is sketched in Figure \ref{scheme_fig}. In this scheme, HXRs are associated with electrons
impacting $z$ = 0, where they are stopped by the rapidly increasing
density\footnote{Strictly speaking, $z$ = 0 delineates the photosphere probed by
the SOHO/MDI magnetogram, and not the density step (transition region) seen by the precipitating
electrons. We shall not make this distinction here since our model does not include a background density profile.}. 
Electrons which remain magnetically trapped are associated with gyrosynchrotron 
emission, and electrons which leave the simulation domain at $z$ = $H$ are associated with radio 
type III emission at the local plasma frequency.

\begin{figure}[ht]
\centerline{\includegraphics[width=0.48\textwidth,height=0.65\textwidth]{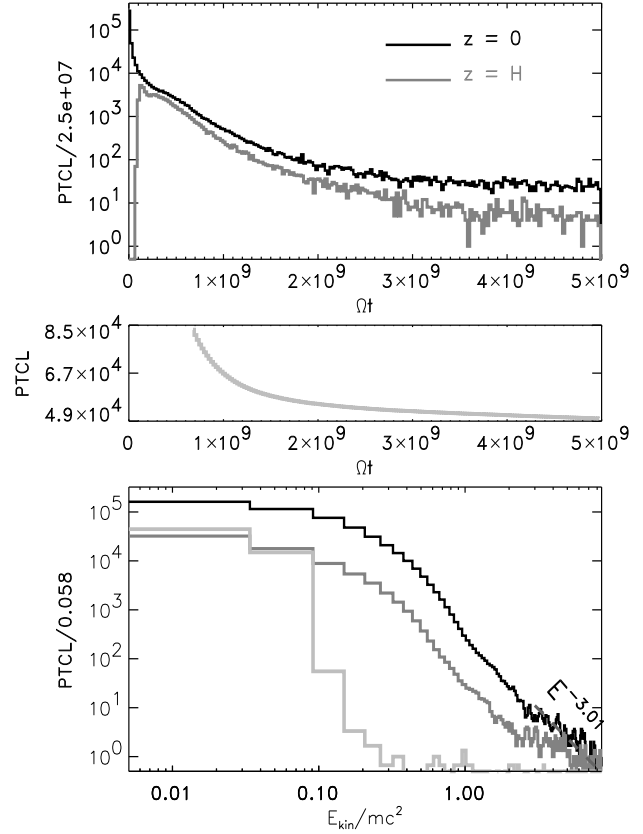}}
\caption{Precipitating, escaping, and trapped electrons. Top: exit rates at the
slab boundaries. Middle: remaining (trapped) particles. Bottom: terminal kinetic energies.
$\Omega t = 5 \cdot 10^9$ corresponds to 5 seconds in real time.}
\label{sim_ptcls_fig}
\end{figure}

\begin{figure}[h]
\centerline{\includegraphics[width=0.45\textwidth,height=0.28\textwidth]{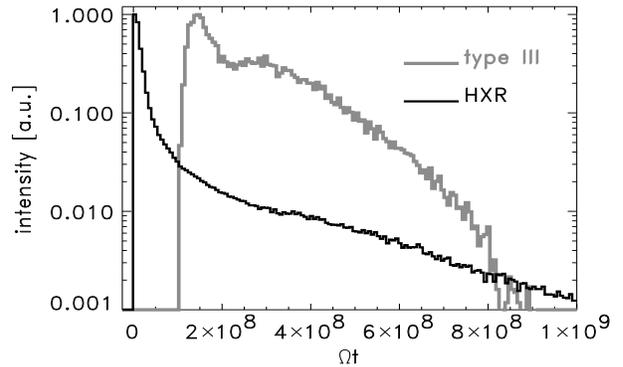}}
\caption{Simulated HXR and type III light curves.}
\label{sim_lightcurves_fig}
\end{figure}


We have performed several simulation runs with varying parameters of the electromagnetic fields. One of these
runs is discussed in detail in Sections \ref{electron_sect} to \ref{gyro_sect}, while the outcome of
the others is summarized in Section \ref{other_sim_sect}. In all simulations, electrons with initial velocity 
$v_0 = 0.1c$ ($E_0 = 2.6$ keV) are injected at $t$ = 0 inside the dissipation regions, and traced until 
one of the following conditions is met: $z < 0$ (`precipitation'), $z > H$ (`escape'), or $t > T_{\rm max}$ 
(`trapping'). The simulation duration $T_{\rm max}$ is long compared to the average arrival times at $z = 0$ 
and $z = H$. $T_{\rm max}$ is also long compared to the free escape time $\tau_0 \doteq H/v_0$ calculated 
on grounds of the initial velocity. The synchronized injection at $t$ = 0 is somewhat artificial since we use
magnetograms of a non-flaring active region. We have chosen this initial condition in order to model
HXR and radio transients and thus provide additional (timing) diagnostics
in situations where an observable flare pulse occurs,
based on the assumption that flares are large-scale versions of the process considered here. The predictions
will, though, be qualitative only, since the time evolution of the flaring region is not taken into account.

\subsection{\label{electron_sect}Electrons}

The outcome of a typical simulation is summarized in Fig. \ref{sim_ptcls_fig}. The top panel represents the exit rates
at $z$ = 0 and $z$ = $H$. Most particles exit through the $z$ = 0 boundary and the 
precipitation rate has a sharp peak at $\Omega t \sim 10^7$ (not resolved in Fig. \ref{sim_ptcls_fig}). 
The escape rate peaks later. The remaining (trapped) population gradually decreases (Fig. \ref{sim_ptcls_fig} middle;
times smaller than the free escape time  $\Omega \tau_0 = 6 \cdot 10^8$ are not shown). 
A detailed look on the particle orbits shows that these usually perform numerous
visits to the dissipation regions before reaching relativistic energies. The fraction of time spent by the 
electrons inside the dissipation regions is quite different for the different populations. While precipitating particles 
spend about 10\% of their time inside the dissipation regions, the escaping ones do so only for about 2\%. The trapped ones
spend 14\% of their time inside the dissipation regions, but do not systematically gain energy. In contrast, the
escaping and -- more pronounced -- precipitating particles gain relativistic energies. The terminal energy spectra
are shown in Fig. \ref{sim_ptcls_fig} (bottom). Non-Maxwellian tails occur in the precipitating spectrum at energies $E/mc^2$ $\sim$ 1. 
The trapped electron spectrum (Fig. \ref{sim_ptcls_fig} light gray) refers to the end of the simulation ($t = T_{\rm max}$). 

\begin{figure}[h]
\centerline{\includegraphics[width=0.48\textwidth,height=0.35\textwidth]{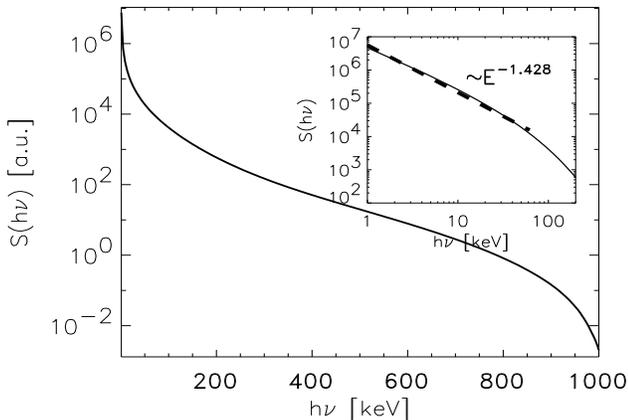}}
\vspace{-5mm}
\caption{Thick-target bremsstrahlung spectrum from the precipitating electrons.}
\label{sim_hxr_spec_fig}
\end{figure}

\subsection{\label{HXR_sect}Hard X rays}
 
\paragraph{Light Curve and Spectrum.}In our model, the precipitating population is associated with HXR bremsstrahlung, with
the HXR time profile being directly proportional to the electron exit time density at $z$ = 0 (Fig. \ref{sim_lightcurves_fig}). 
Since $z$ = 0 is an absorbing boundary we use the thick-target approximation. The
(orientation-averaged) bremsstrahlung cross section is taken from the fully relativistic 
formula 3BN of Koch \& Motz (\cite{koch59}). The resulting HXR spectrum is shown in Figure \ref{sim_hxr_spec_fig}.
At energies up to $\sim$40 keV, it can be fitted by a (hard) power law with index 1.4 (Fig. \ref{sim_hxr_spec_fig} inlet). At higher
energies, it decays exponentially and finally drops off faster than exponentially at $\sim$800 keV. Most of the exponential part and
the super-exponential decay could usually not be observed with X-ray observatories like RHESSI because of the limited
dynamic range (2-3 orders of magnitude for M class flares; see Grigis \& Benz \cite{grigis04}).

\begin{figure}[h]
\centerline{\includegraphics[width=0.48\textwidth,height=0.27\textwidth]{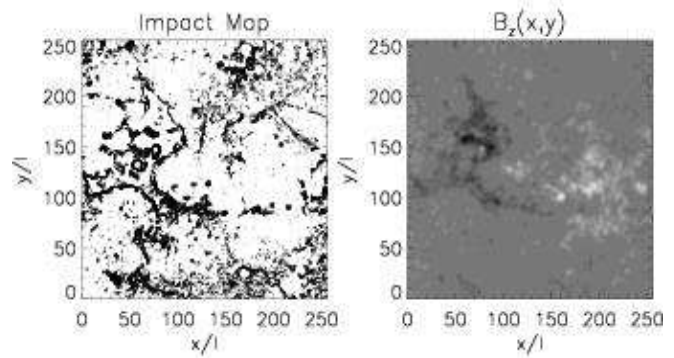}}
\caption{Electron precipitation sites (left) and boundary magnetic field (right). Positive
magnetic polarity is light; negative polarity is dark.}
\label{impact_fig}
\end{figure}

\paragraph{Impact Map.}The present simulation also predicts the sites at which electrons impinge onto
the chromosphere (Fig. \ref{impact_fig} left), and allows {\it a comparison of this} with the boundary magnetic field
(Fig. \ref{impact_fig} right, repeated here from Fig. \ref{b0_fig} for better clarity). Since we inject particles 
in all dissipation regions simultaneously, the impact map represents a statistical prediction. There are two observations 
to be made: first, the electrons preferentially precipitate at negative magnetic polarity. Secondly, 
the impact density peaks not at the center of the sunspots or filaments, but at their borders. The first observation is
a simple consequence of the constant-$\alpha$ assumption, by which the electric acceleration is always
parallel or antiparallel to the magnetic field\footnote{We have assumed here positive $\alpha$ and that
the electrons have positive charge.}. The second
observation is a consequence of the mirroring in converging magnetic field lines, which makes it hard for electrons
to penetrate down to $z$ = 0 in regions of maximal $|{\bf B}(x,y,0)|$. The actual
impact map represents a trade-off between the number of downstreaming electrons and their reflection probability,
which is optimal at the sunspot/filament boundaries.

\begin{figure}
\centering
\includegraphics[width=0.43\textwidth]{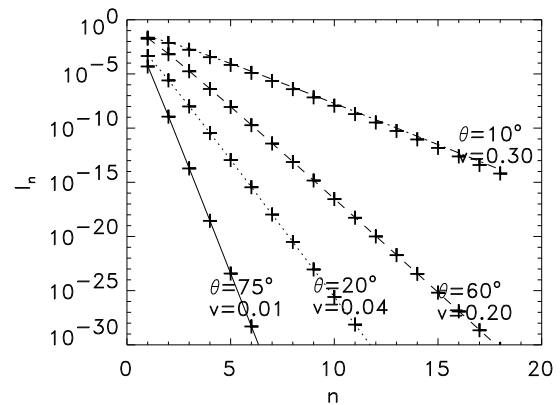}
\caption{Gyrosynchrotron intensity according to the exact expression 
(Eq. \protect\ref{schott} -- crosses) and its approximation (Eq. \protect\ref{schott_approx} -- dotted line).}
\label{schott_approx_fig}
\end{figure}

\begin{figure}
\centering
\includegraphics[width=0.45\textwidth]{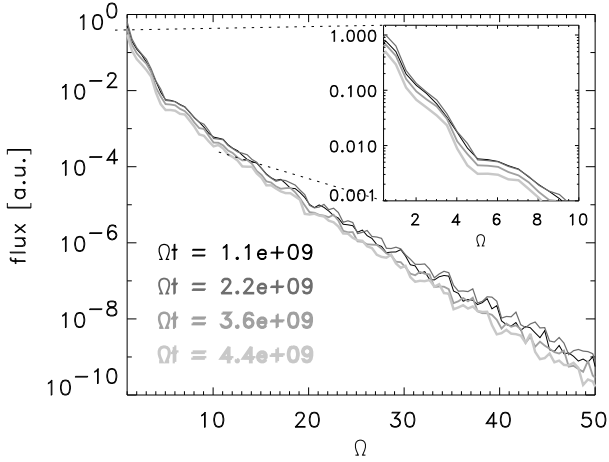}
\caption{Optically thin gyrosynchrotron spectrum emitted in $z$ direction.
Different curves represent different times. $\Omega = 1$ corresponds to about 1 GHz.}
\label{sim_gyro_spec_fig}
\end{figure}

\subsection{\label{beam_sect}Electron beams}

Electrons which escape to altitudes $z$ $>$ $H$ are considered as a proxy for type III radio bursts. Contrary to the
incoherent bremsstrahlung and gyrosynchrotron radiation, the plasma emission of type III bursts is coherent and
therefore not proportional to the number of emitting electrons. The non-linear dependence (say, $\propto N^\xi$ with 
$\xi > 1$) compensates for the 
relatively tenuous escaping population. The resulting time signal is shown in 
Fig. \ref{sim_lightcurves_fig}, together with the HXR light curve. The scaling of the two
intensities is arbitrary. As the background plasma density is not specified in our model, we cannot
predict the emission frequency of the plasma waves. However, we may consult observed density
profiles at height $H \sim 3 \cdot 10^4$ km, yielding $n_e \sim 2.5 \cdot 10^8$ cm$^{-3}$ 
averaged over the quiet corona (Fludra \cite{fludra99}). This number varies by
a factor 3 when equator-to-pole variation is taken into account (Gallagher et al. \cite{gallagher99}), 
and may be larger by a factor 10 above faculae (Dumont et al. \cite{dumont82}). Thus we 
associate a plasma density of 10$^8$ cm$^{-3}$ to 2$\cdot$10$^9$ cm$^{-3}$ 
with the $z$ = $H$ boundary of our simulation slab, corresponding to a plasma
frequency of 100 to 400 MHz.

\begin{figure*}[ht]
\centering
\includegraphics[width=0.82\textwidth,height=0.85\textwidth]{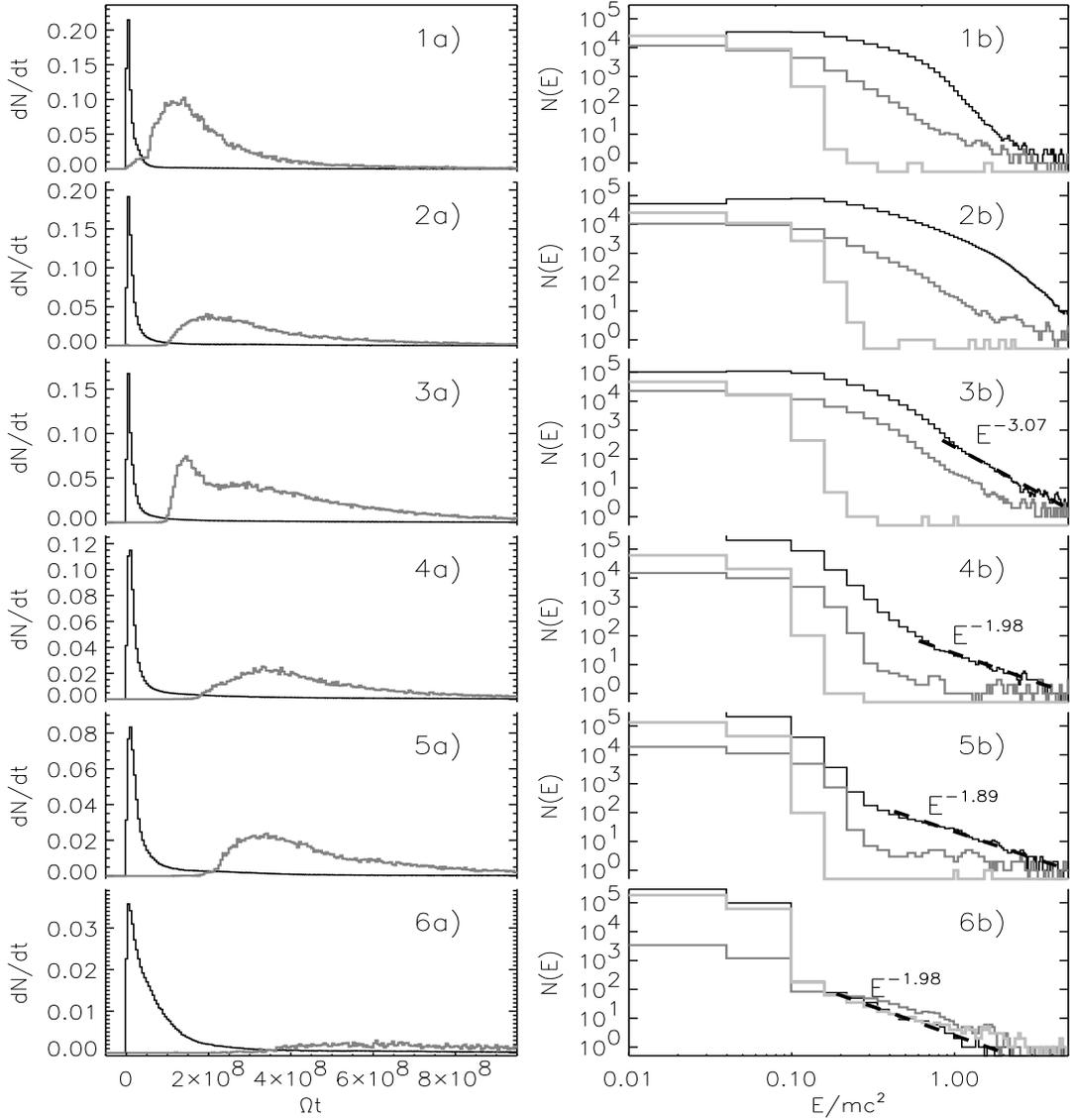}
\caption{Parameter exploration. Left column: exit rates. 
Right column: terminal energy spectra. Black (gray, light gray) lines refers 
to precipitating (escaping, trapped) populations. See Tab. \protect\ref{sim_tab}
for the simulation parameters.}
\label{overview_fig}
\end{figure*}

\subsection{\label{gyro_sect}Gyrosynchrotron emission}

The electrons which are trapped in closed loops are thought to be responsible for
gyrosynchrotron radiation, manifesting in microwave solar continuum bursts\footnote{In contrast,
narrowband decimetric continua are better explained by transition radiation (Fleishman et al. \cite{fleishman05},
Nita et al. \protect\cite{nita05}).}. 
In this case there are many bursts which are optically thin at all frequencies (Fleishman et al. \cite{fleishman03}), 
and we assume here optically thin radiation for simplicity. Moreover, we neglect the plasma response
and consider emission in vacuum. Therefore, our results apply to the emission at frequencies above the spectral 
peak provided by either optical thickness or the Razin effect. An electron on a circular orbit emits then at 
frequency $n \omega / \gamma$ the intensity (Schott, \cite{schott12})

\begin{equation}
I_n = \frac{n^2 \omega^2}{\gamma^2} \left( \tan^2 \theta \; J_n^2 (nv\cos \theta) + v^2 {J'_n}^2 (nv\cos \theta) \right)
\label{schott}
\end{equation}

where $n$ is an integer, $\omega$ = $eB/m$ is the local cyclotron frequency, and $\theta$ is the angle between 
the gyration plane and the line of sight, which latter is taken along the $z$-direction. 
In our simulation, most trapped electrons have velocities $v$ $\ll$ 1 (Fig. \ref{sim_ptcls_fig} bottom), so that 
the argument of the Bessel functions in Eq. (\ref{schott}) is small and $J_n(z)$ can be approximated
by $(2\pi n)^{-1/2}(ez/2)^n$ (Abramowitz \& Stegun, \cite{abramowitz70}), yielding

\begin{equation}
I_n \simeq \frac{\omega^2}{\gamma^2} \frac{1 + \sin^2 \theta}{\cos^2 \theta} \frac{n}{2 \pi} 
\Big| \frac{e v \cos \theta}{2} \Big|^{2n} \, ,
\label{schott_approx}
\end{equation}

where Euler's number $e$ = 2.718 is not to be confused with the elementary charge. The relative accuracy
of the approximation (\ref{schott_approx}) is of order $|v \cos \theta|$ for all $n$;
a numerical illustration is shown in Fig. \ref{schott_approx_fig}. 
From Eq. (\ref{schott_approx}) we see that the intensity $I_n$ decays exponentially 
with $n$, so that the radiation is concentrated at low harmonics. In the limits $v \to 0$ and $\theta \to 90^o$ 
only the fundamental ($n$ = 1) contributes. If the velocity has a component parallel to the magnetic field
and to the observing direction, this results in a Doppler shift $\omega \to \omega^*$, and
$v$ in Eqns. (\ref{schott}, \ref{schott_approx}) has to be replaced by $v_\perp$, the component perpendicular 
to the magnetic field. Neglecting dispersion and absorption, 
we may thus obtain the radio spectrum by accumulating a histogram of the frequencies $n \omega^* / \gamma$ 
with weight $I_n$ for all particles and times. The result is shown in Figure 
\ref{sim_gyro_spec_fig}. Note that the spectrum decays monotonically and approximately exponentially with
frequency, which is different from the familiar synchrotron ($\gamma \gg 1$) shape with powerlaw
rise and -decay. Here, the exponential decay is mostly caused by the harmonics 
dependence (Eq. \ref{schott_approx}) but also supported by the exponential decay of magnetic field with height.
Individual harmonics are washed out because of the magnetic field inhomogeneity.
\begin{table}[h]
\begin{tabular}{cccccccc}
run &  $N_k$ & $N_x$ & $l$ [km] & $H/l$ & $\alpha l$ & $\eta$ [c$^2$/$\Omega$] & $h_d/l$ \\\hline
1) & 5000 & 128 & 875 & 40 & 0.05 & 4.38 & 0.56 \\
2) & 5000 & 256 & 660 & 50 & 0.05 & 6.84 & 0.50 \\
3) & 5000 & 256 & 660 & 50 & 0.03 & 6.84 & 0.19 \\
4) & 5000 & 256 & 660 & 64 & 0.01 & 6.84 & 0.13 \\
5) & 1000 & 256 & 660 & 64 & 0.01 & 6.84 & 0.05 \\
6) & 1000 & 256 & 660 & 64 & 0.01 & 1.37 & 0.08 \\
\end{tabular}
\caption{Simulation parameters. In all simulations, $\Omega^{-1} = 10^{-9}$s and $u_c l$ = 0.2.}
\label{sim_tab}
\end{table}

\subsection{\label{other_sim_sect}Parameter exploration}

We have performed various production runs, involving a total CPU time of about 4 months. Different runs have different 
values of the parameters $H$ (slab height), $\alpha$ (force-free parameter), and $\eta$; see Table \ref{sim_tab}. 
The critical twist is fixed at $u_c l = 0.2$. The simulation discussed in Sections \protect\ref{electron_sect} 
- \protect\ref{gyro_sect} is labeled 3 and represents an intermediate case. 
Each simulation involves several $10^5$ particles. An overview on the simulation results is presented in Fig. 
\ref{overview_fig}. The column a) contains the electron exit rates at $z$ = 0 (black line) and
$z$ = $H$ (gray line) during the first third of the simulation time, normalized by the total number of
simulated electrons. The column b) shows the energy distributions
at the end of the simulation ($t$ = exit time, or $t = T_{\rm max}$). 
The translation of the electron results onto HXR and radio wave predictions proceeds similar
as in Sections \ref{HXR_sect} to \ref{gyro_sect}, and preserves the hardness ordering of the spectra.

\begin{figure*}[ht]
\centering
\includegraphics[width=0.85\textwidth]{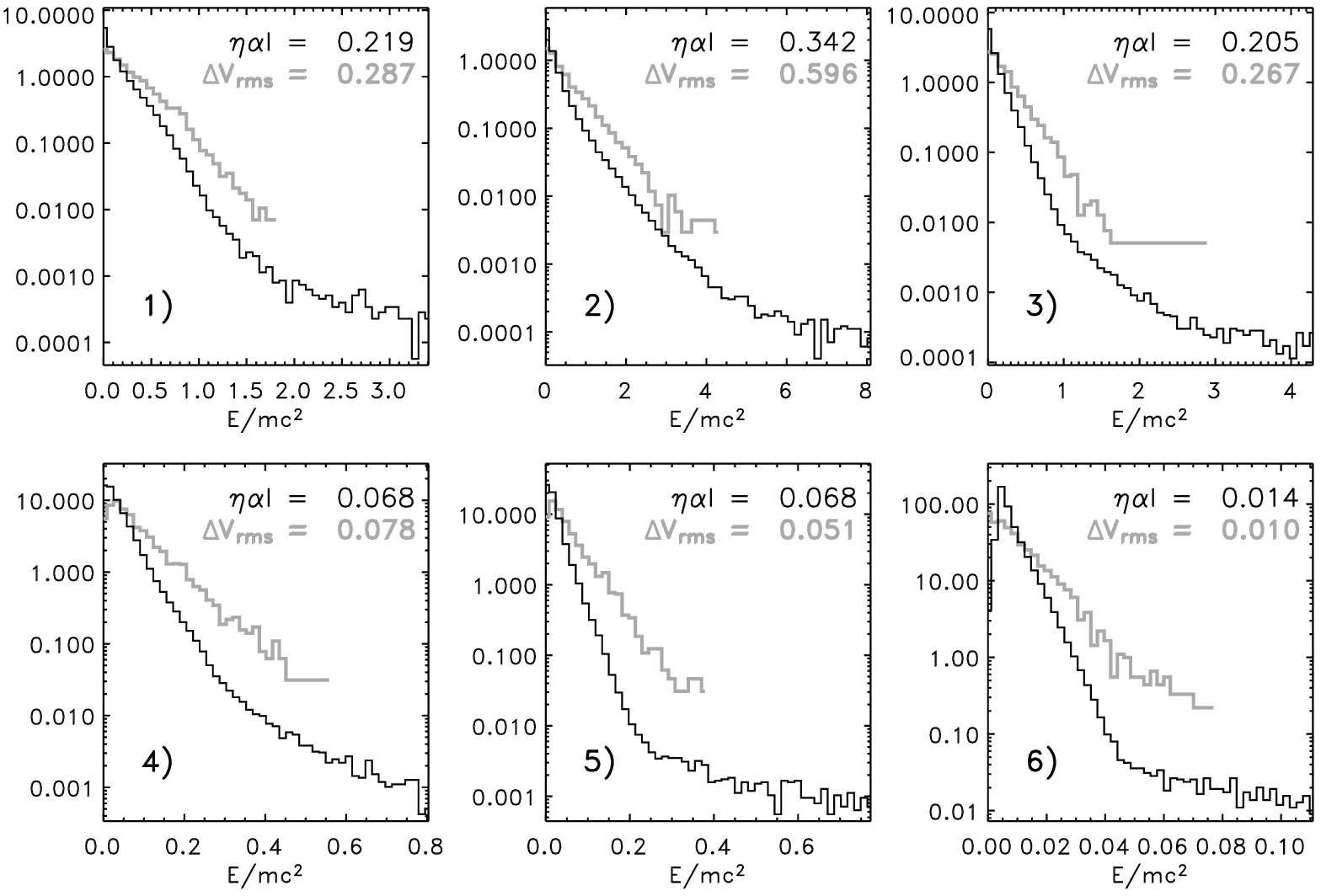}
\caption{The distribution of precipitating, escaping, plus trapped particle energy 
(black line) and the distribution of voltage drops (gray line).}
\label{potentialdrop_fig}
\end{figure*}

The simulations 1 to 6 may be summarized by saying that the results of Sections 
\ref{electron_sect} to \ref{gyro_sect} are generic, in the sense that the general shape and relative 
timing of the light curves are robust against variation of ($H, \alpha, \eta$). However, the time scaling, 
the division into precipitating and escaping populations, and the spectrum
depend on the parameters (Tab. \ref{sim_tab}). The time delay between the precipitation
pulse and the onset of escape (Fig. \ref{overview_fig} left) increases, naturally, with the 
slab height $H$. It also depends on the force-free parameter $\alpha$, yet in a more subtle way. 
From the comparison of simulations 2 and 3 it is seen that a larger $\alpha$ favours prompt escape. This is not 
immediately obvious, since a larger $\alpha$ implies a smaller force-free scale $1/\alpha$ and thus
smaller magnetic loops which do not protrude to $z$ = $H$. However, a larger $\alpha$ also admits more 
non-decaying modes ($k_\perp < \alpha$) and thus more open field lines along
which the electrons can escape. Moreover, a larger $\alpha$ implies larger curvature, so that the criticality 
condition (Eq. \ref{crit_cond}) is more frequently met and the accelerating volume is larger
($h_d$ = 0.5$l$ in simulation 2 but $h_d$ = 0.19$l$ in simulation 3). As a net effect, a large force-free parameter 
$\alpha$ thus favours rapid acceleration and escape. A similar enhancement of acceleration by the presence of 
small scales accounts for the difference between panels 4 and 5 in Fig. \ref{overview_fig}. These simulations
differ only by the number of Fourier components. 
Run 4 contains wave vectors up to $k_\perp l \le$ 1.94, whereas run 5 contains wavevectors $k_\perp l \le$ 0.8.
Accordingly, run 4 reaches somewhat higher energies (panels 4b and 5b). The escape time also 
increases (and the escape probability decreases) with decreasing anomalous resistivity and thus with
decreasing electric field (5a vs. 6a).
The trapped component (light gray) is throughout softest, except for the case of very small resistivity (run 6) where 
all 3 populations behave similarly for $E/mc^2 \ga 0.1$. Run 6 is similar to run 5 but with
smaller resistivity, compensated by a somewhat larger dissipation volume.

As a general trend, it can be seen from Fig. \ref{overview_fig} (top to bottom) that smaller 
dissipation volumes and weaker electric fields yield a more gradual evolution and less intermediate
energies (0.1 $< E/mc^2<$ 1) but a comparable amount of high energies, so that the energy histograms have more 
tenuous but harder tails. This reflects a change of the nature of acceleration from frequent and small 
energy increments to rare but violent energy gains, in the course of which the stochastic process 
leaves the domain of attraction of the central limit theorem.

\section{Summary and Discussion}

We have simulated gyrokinetic electron orbits in constant-$\alpha$ force-free magnetic fields with
anomalous resistivity $\eta$. The latter is localized in (postulated) dissipation regions where the magnetic twist 
$\nabla \times \bhat$ exceeds a given threshold $u_c$, corresponding to a critical scale $1/u_c$ $\sim$ 3000 km.
The dissipation regions cover about $10^{-3}$ of the simulated slab volume, and can be characterized by a 
column height $h_d$ of 30 to 300 km. The (parallel) dissipative electric field exceeds the Dreicer field by one or 
two orders of magnitude and yields direct acceleration of runaway electrons.
In general, the electrons visit numerous dissipation regions before reaching relativistic energies,
and arrive at the chromosphere before escaping to the higher corona. The latter ordering agrees
with an observed trend for HXR to precede type III bursts (Aschwanden et al. \cite{aschwanden92}; 
Arzner \& Benz \cite{arzner05a}), also in terms of absolute time delays.

From a physics point of view, one may compare the energies reached in Fig. \ref{overview_fig} to
the voltage drop along the magnetic field line inside a (simply connected) dissipation region ${\cal D}$.
(We use here the term `voltage drop' rather than `potential drop' since ${\bf E} = \eta {\bf j}$ is
not a potential field.) The quantity $\Delta V = \int_{\cal D}{\bf E} \cdot d {\bf l}$ gives an upper limit 
on the energy which can be gained inside a single dissipation region; reflection from converging
field lines (and also deceleration from ${\bf E}$ itself)
does generally prevent the particles from exploiting the full voltage drop.
By choosing 5$\cdot$10$^3$ random points inside the dissipation regions, and following
the magnetic field lines going through these points, we find
the distribution of voltage drops shown in Fig.\ref{potentialdrop_fig} (gray line), together 
with the distribution of all terminal electron energies (black line). Both kinetic energy and voltage 
drops are measured in units of the electron rest mass. As can be seen, the two distributions
scale similarly in energy from one simulation to another (note the different energy axes!),
and the sub-exponential tails of the particle energy occur above the largest voltage drop 
present in the simulation. This agrees with the observation that many dissipation regions are visited
before the particles reach the highest energies. The root mean square voltage drops $\Delta V_{\rm rms}$
are of the order of the voltage drop across the magnetogram resolution, $\eta \alpha l$.
The distribution of voltage drops decays slower than the distribution of particle
energies, which relates to the fact that $\Delta V_{\rm rms}$ is merely an upper bound on 
the kinetic energy gain. The peak of the particle distributions at lowest energies
is mostly due to the trapped component.

The present model predicts a statistical preference for the HXR to occur at one magnetic 
polarity, because of the assumption of constant $\alpha$, by which the electric field ${\bf E} = \eta \alpha {\bf B}$ 
is always parallel ($\alpha > 0$) or antiparallel ($\alpha < 0$) to the magnetic field. The electrons preferentially
impact the chromosphere not in the center of sunspots or filaments (where $|{\bf B}|$ is largest) but at 
their borders. Ions, due to their opposite charge, should preferentially impact at opposed magnetic polarity
(but the ion dynamics is not directly comparable to the electron dynamics because of the large mass ratio).
The experimental verification or falsification of this model prediction, using non-randomized force-free extrapolations
and spatially resolved X-ray (e.g., RHESSI) observations of flaring active regions, will be the subject of 
future investigations. It should also be pointed out that the impact map (Fig. \ref{impact_fig}) is obtained 
from simultaneous injection in all dissipation regions. If injection was restricted to a single dissipation region 
or group of dissipation regions, then only few impact regions would occur, as usually observed in individual solar flares.

Finally, we should mention that the present approach has many caveats. Most prominent among them is the fact that
test particle simulations can not tell us anything about the absolute number of electrons which are accelerated, and
their application to the real solar corona requires caution in order to match global electrodynamic constraints 
(i.e. return currents; Spicer \& Sudan \cite{spicer84}). As a rule, our simulations can only account for a tenuous 
high-energy tail, because the back-action on the electromagnetic field and Coulomb collisions are neglected.
Then, we have used here magnetograms of non-flaring active regions -- where the force-free
extrapolation should be a good approximation -- and have thus envisaged `quiescent' coronal heating by nanoflares
rather than large isolated events. The observable predictions, though, address sizable flares where
all types of emission can be detected. We suggest that the actual flares are generated by 
a similar process, and should thus have similar characteristics. 
Conversely, the non-flaring active regions are constantly doing what 
flaring active regions do but in smaller, possibly undetected numbers. This coronal process may account for 
spatial X-ray and radio fine structures in quiet solar regions (Benz et al. \cite{benz97}) and non-thermal
electrons measured in space during quiet times (`superhalo'; Lin \cite{lin98}), as an alternative explanation 
of MHD wave acceleration in the solar wind. Also, the assumption of a time-independent magnetic field is an
approximation only. It is certainly violated during large flares with re-structuring of the global magnetic
topology. Also, it does not formally account for the continuous footpoint motion driving the Parker 
(\cite{parker83}) scenario. However, it may be a reasonable approximation for medium-size flares, where the
Parker mechanism does not destroy the overall magnetic field structure. This point of view is supported
by the work of Aulanier et al. (\cite{aulanier05a,aulanier05b}) who numerically study the evolution of magnetic
flux tubes and quadrupolar configurations under sub-Alfv\'enic photospheric motion. Their (incompressible, resistive) 
MHD simulations demonstrate that current sheet formation and topological changes may be described by a sequence of
quasi-equilibria, so that a static approximation over a few ten seconds seems appropriate.
With regard to particle acceleration, the temporal evolution 
of the magnetic fields is expected to decrease trapping and facilitate both precipitation and escape.

\begin{acknowledgements}
The authors thank Emmanuela Rantsiou and Manolis Georgoulis for help with SOHO/MDI magnetograms, 
and Arnold Benz and Gregory Fleishman for helpful discussions and comments. This 
work was supported in part by the Research Training Network (RTN) `Theory, Observation and 
Simulation of Turbulence in Space Plasmas', funded by the European 
Commission (contract No. HPRN-eT-2001-00310).
\end{acknowledgements}   

\begin{figure}[ht]
\centering
\includegraphics[width=0.5\textwidth]{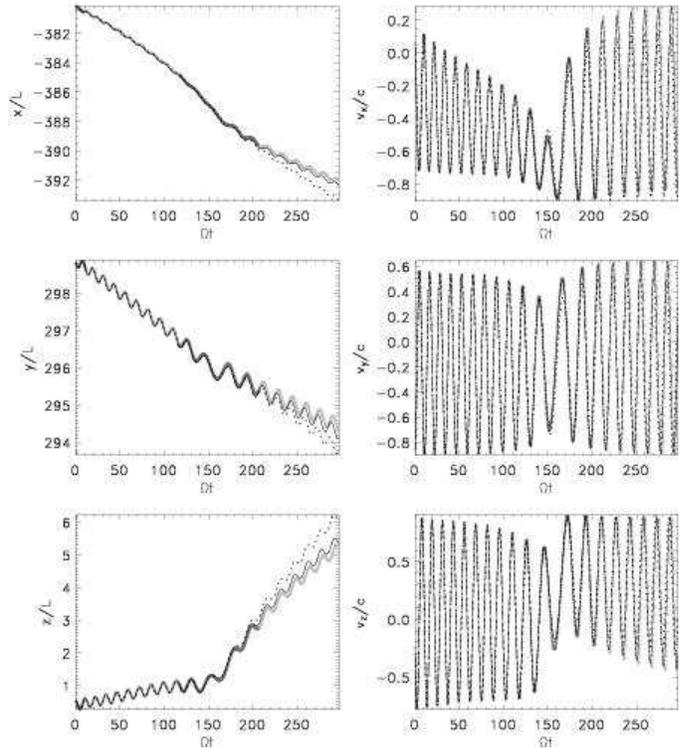}
\caption{Benchmark of the orbit integration schemes, showing position (left column) and velocity 
(right column) of a relativistic electron in strongly curved ($l/l_0$=10) magnetic fields. Solid line represents
exact orbits (Eqns. \protect\ref{dxdt},\protect\ref{dvdt}); dotted line, gyrokinetic approximation
(Eqns. \protect\ref{dXdt} - \protect\ref{duds}). Gray crosses represent the hybrid mode, with
the exact regime marked dark and the gyrokinetic regime marked light.}
\label{orbits_fig}
\end{figure}

\appendix

\section{\label{numerics_appendix}Numerical Implementation}

\subsection{Regime switching and accuracy control}

The choice of the orbit integration method, exact or gyrokinetic, depends on two constraints. 
From a physical point of view, the gyrokinetic approximation requires the Larmor radius 
to be small compared to the scale of the magnetic field, $\rho \ll l$. From a numerical point of view, the 
gyrokinetic approximation is faster than exact orbit integration whenever the latter is dominated by particle 
motion rather than by field evaluation. Indeed, if $\tau_f$ denotes the CPU time needed for a single field 
evaluation and $\tau_g$ is the CPU time needed to resolve a single gyration, then the gyrokinetic approximation 
is by a factor

\begin{equation}
f = \frac{\tau_g}{\tau_f} \times \frac{l}{\rho}
\label{benefit}
\end{equation}

faster than exact orbit tracking. The time $\tau_g$ is approximately independent of the 
integration scheme as long as this is optimal, and a typical value on a medium-size workstation is found to be
$\tau_g \sim 5 \cdot 10^{-5}s$. The time cost of a single field evaluation, on the same hardware, is 
$\tau_f \sim N_k \times 5 \cdot 10^{-6}$s with $N_k$ the number of Fourier components. Therefore the
gyrokinetic method is numerically beneficial for $N_k < 10 \times l/\rho$. Thus, if the 
gyrokinetic approximation is physically allowed ($l/\rho \gg 1$) then it is also numerically
beneficial in the sparse-Fourier representation; the upper bound on $N_k$ is in
practice rarely met.

We turn now to the precise (and implemented) formulation of regime switching and accuracy control. The gyrokinetic
approximation requires that the magnetic field should change slowly across a Larmor radius,
$(\nabla {\bf B}) \cdot \mrho < \epsilon_g B$, where $\nabla {\bf B}$ is the Jacobian and $\epsilon_g \ll 1$.
For practical purposes, we replace this by the stronger and computationally less expensive constraint
$\rho^2 \sum_{ij}|\partial_i B_j|^2 < \epsilon_g^2 B^2$ with $\rho = |{\bf p}|/B$. The quantity $\rho$
also includes parallel momentum, so that the condition accounts for particle acceleration inside the dissipation regions.
In addition to the gyrokinetic approximation error, there are numerical errors from
the finite time step and field computation rate. These are controlled by limiting the
Cash-Karp position- and momentum errors, and by enforcing re-calculation of the electromagnetic fields if
$|\Delta {\bf x}|^2 \sum_{ij} |\partial_i B_j|^2 > \epsilon_B^2$,
where $\Delta {\bf x} = {\bf x} - {\bf x}_{old}$ is the distance to the last field evaluation 
point ${\bf x}_{old}$. Using the same Jacobian, the fields are linearly extrapolated from
${\bf x}_{old}$. In order to avoid relentless switching between exact and gyrokinetic regimes,
a minimum duration of one gyro period in each of the regimes is enforced.

\subsection{Benchmarks}

The exact orbits have been computed with different integration- and field evaluation schemes.
The time integrators include the Verlet scheme and other leapfrog variants, and Runge-Kutta 
schemes with and without adaptive time stepping. The field evaluation was done either pointwise or with linear 
interpolation involving $\nabla {\bf B}$. Once the exact orbits were established, they have been
used to benchmark the gyrokinetic orbits. As an example, Figure \ref{orbits_fig} displays 
individual cartesian components of position and velocity in an extreme situation, i.e. for a 
highly relativistic ($v$ = 0.99$c$) particle with small magnetic scales ($l/l_0$ = 10). Solid line represents
exact orbits, dotted line represents the gyrokinetic approximation, and gray represents the hybrid mode with
automatic regime switching, where exact (gyrokinetic) regimes are marked dark (light).
In the hybrid method, the particle starts with exact orbit integration and switches to the gyrokinetic
description at $\Omega t$ $\sim$ 10. It remains then in the gyrokinetic mode until at $\Omega t$ $\sim$ 120 a
region of strong magnetic curvature is encountered, where exact orbit tracing is enforced (dark gray). 
At $\Omega t$ $\sim$ 200 the orbit switches back to the gyrokinetic method (light gray). During the
gyrokinetic phase, the magnetic moment is conserved to within 0.1\%, but varies during the exact-orbit
phase by some 80\%. As can be seen, the hybrid orbit is closer to the exact result than the purely gyrokinetic orbit.

\end{document}